# Irradiation-induced metal-insulator transition in monolayer graphene

I. Shlimak*, E. Zion, A. Butenko, Yu. Kaganovskii, V. Richter, A. Sharoni, E. Kogan and M. Kaveh

*Jack and Pearl Resnick Institute of Advanced Technology, Department of Physics, Institute of Nanotechnology and Advanced Materials Bar Ilan University, Ramat Gan 52900, Israel*

*Corresponding author. Email address: shlimai@biu.ac.il

**Abstract**

A brief review of experiments directed to study a gradual localization of charge carriers and metal-insulator transition in samples of disordered monolayer graphene is presented. Disorder was induced by irradiation with different doses of heavy and light ions. Degree of disorder was controlled by measurements of the Raman scattering spectra. The temperature dependences of conductivity and magnetoresistance (MR) showed that at low disorder, conductivity is governed by the weak localization and antilocalization regime. Further increase of disorder leads to strong localization of charge carriers, when the conductivity is described by the variable-range-hopping (VRH) mechanism. It was observed that MR in the VRH regime is negative in perpendicular fields and is positive in parallel magnetic fields which allowed to reveal different mechanisms of hopping MR. Theoretical analysis is in a good agreement with experimental data.

# I. Introduction

The main purpose of the experiments presented in this review article was to study the transformation of electron transport in monolayer graphene (MG) induced by increasing of disorder. It was expected to observe the gradual localization of the charge carriers, up to the metal-insulator transition and transformation of the mechanism of conductivity from metallic one to hopping, which is characterized for strongly localized charge carriers.

MG is a gapless semimetal with a conical linear electronic dispersion (see pioneering work [1] and some reviews [2-4]). The conduction band (CB) formed from $\pi^*$-states and valence band (VB), formed from $\pi$-states touch in so-called Dirac points (DP), Fig. 1. There are six valleys of two non-equivalent kinds $K$ and $K'$. The charge carriers - electrons and holes near the DP are massless and known as "Dirac fermions". Pristine MG is always conductive: even in the case when the Fermi level is located in the DP between the CB and VB, conductivity is non-zero and has a minimum value $\sigma_{min}$. Usually, $\sigma_{min} \approx 4e^2/h$, ($e^2/h = 38.7$ $\mu$S is the quantum of two-dimensional conductivity), but other values were also observed [2]. Any shift of the Fermi energy above or below the DP leads to an increase in the concentration of electrons or holes and to an increase in conductivity (Fig. 2). The value $\sigma_{min}$ corresponds to a maximum resistivity of the order of 6-10 k$\Omega$ per square. Meanwhile, much larger resistivity is required for different applications, for example, to increase the ON-OFF ratio in the field effect transistors, or to separate electrically the microcircuits fabricated on the common surface, etc.

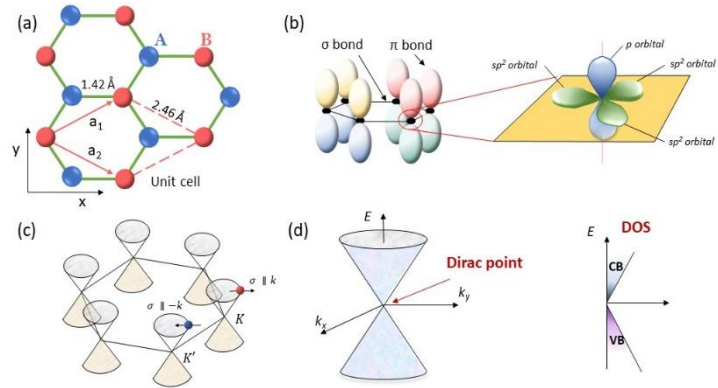

**Figure 1.** Schematics of the band structure of monolayer graphene. (a) honeycomb lattice with two sub-lattices (A and B); (b) $\sigma$- and $\pi$-bonds and sp$^2$-orbitals; (c) K- and K'- valleys with "pseudospin" parallel (anti-parallel) to the momentum; (d) conical dispersion law

$E = \hbar v_F |\mathbf{k}|$, where $v_F \approx 10^6$ m/s is the Fermi velocity, and (e) linear density of states (DOS) in conductive band (CB) and valence band (VB).

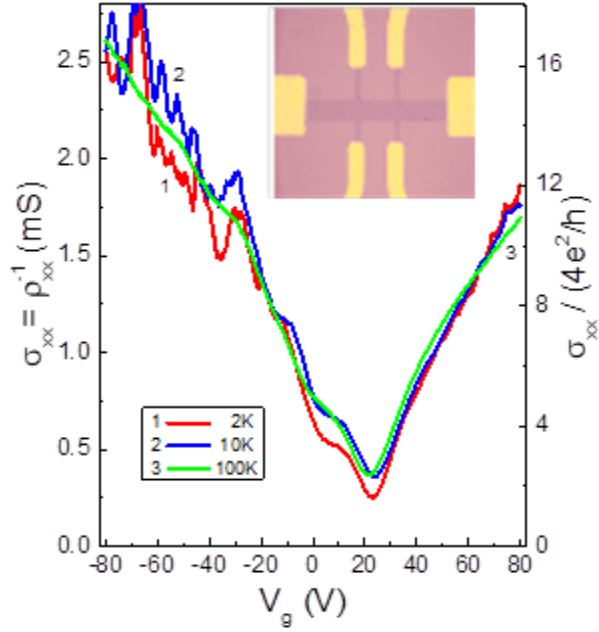

**Figure 2.** Longitudinal conductivity $\sigma_{xx} = (\rho_{xx})^{-1}$ of non-irradiated MG sample as a function of the gate voltage at different temperatures. Conductivity is shown in mS (left scale) and in dimensionless units $\sigma/(4e^2/h)$ (right scale). Inset shows the optical image of the sample, the sample width is 10 μm.

In semiconductor systems, an increase in the resistance is usually achieved by introducing disorder that leads first to a decrease in mobility due to an increase in scattering, and then to the localization of charge carriers [5]. In the case of MG, disorder was introduced, in particular, by variation of the fabrication conditions [6], by oxidation [7], hydrogenation [8], chemical doping [9], and irradiation by different ions with different energies [10-13]. In distinct graphene-based samples, high resistive state and hopping conductivity was observed earlier [11,14,15]. However, gradual change in the conduction mechanism with increasing disorder was studied insufficiently which gave a motivation for our work. Among different methods to introduce controlled disorder, ion irradiation was chosen because of high accuracy and reproducibility.

## II. Samples and irradiation

There are several methods for obtaining MG samples. The quality of samples is determined by the mobility of the charge carriers. The highest mobility at room temperature was observed in monocrystalline flakes obtained by exfoliation from the bulk graphite, similar to the method used in Ref. [1]. However, these samples are single and they have a small size of about a few micrometers. The larger size samples are usually obtained by molecular epitaxial deposition or by chemical vapor decomposition (CVD) of methane on a metallic catalyst with subsequent transfer to a substrate [16]. Such samples are polycrystalline, have smaller mobility, but they have a size of about a few centimeters, and can be scaled to the larger size.

In our experiments [17-19], large-sized MG specimens 10x10 mm and 5x5 mm were supplied by the "Graphenea" Co. In accordance with certificate, monolayer graphene was produced by CVD method on copper catalyst and transferred to a 300 nm $SiO_2$/Si substrate using a wet transfer process. P-type Si in the substrate was heavily doped with a very small resistivity 0.005 Ohm·cm, which allowed it to be used as a metal gate in the field-effect transistor geometry. The 300 nm thick $SiO_2$ dielectric layer made it possible to observe the graphene samples in an optical microscope as a blue strip (see inset in Fig. 2). The initial specimen of such a large size was not a monocrystalline. It looks like a polycrystalline film with the average size of microcrystals of about a 10 microns. On the surface of these specimens, small micro-samples were fabricated by means of electron-beam lithography (EBL). The sample width was 10 μm, the length – 100 μm, the distance between voltage probes – 30 μm. Deposition of electrical contacts (5 nm Ti and 45 nm Pd) was done by electron-beam evaporation in room temperature and in a high vacuum chamber. Optical observation and electrical testing after EBL showed that some of small samples are damaged; the reduced yield can be explained among other things, by the influence of the triple lift-off process in the case of polycrystalline film. For measurements, only intact samples were selected.

All micro-samples on the surface of each large-scale specimen were grouped into several groups. One group was not irradiated. Samples from this group have a mark "0". Other groups were irradiated with different doses $D$, of light $C^+$ ions (ion mass $M = 12$ in atomic mass units, amu) with energy 35 keV and heavy $Xe^+$-ions ($M = 182$ amu) with the same energy 35 keV. Irradiation was performed on HVEE-350 Implanter.

**III. Determination of the density of structural defects**

To control the degree of disorder of irradiated MG samples, the Raman scattering spectra (RS) were measured in all samples. RS is considered to be an effective tool for probing the structure of disordered graphene films and density of introduced defects [20-22].

Typical RS spectra for disordered graphene consist of three main spectral lines. The G-line at 1600 cm$^{-1}$ is common for different carbon-based materials, including carbon nanotubes, mono- and multilayered graphene, and graphite and is usually used for normalization of the RS.

The 2D-line at 2700 cm$^{-1}$ is related to an inter-valley two-phonon mode, it corresponds to momentum conservation and is emitted in the intact graphene crystalline structure away from any defects. By contrast, the "defect-produced" D-line at 1350 cm$^{-1}$ is related to the inter-valley single phonon scattering process, which is forbidden in the perfect graphene lattice due to momentum conservation, but is possible in the vicinity of a structural defects (edges, vacancies, etc.) where the lattice symmetry is broken. Therefore, the intensity of the D-line in the form of the dimensionless ratio of the amplitudes of the D- and G-lines, $\alpha = I_D/I_G$ is used as a measure of disorder in graphene layers. Conversely, the normalized intensity of the 2D-line, $\beta = I_{2D}/I_G$, is used as a measure of the intact part of the lattice. In disordered samples, in addition to the D-line, a few minor lines connected with different modes of phonons can be seen: D'-line (1620 cm$^{-1}$), D+D'- (2970 cm$^{-1}$), D+D''-line (2450 cm$^{-1}$). The positions of all lines are given for 532 nm excitation laser used in our measurements.

Figure 3 shows the RS spectra for two series of samples irradiated by C$^+$ and Xe$^+$ ions. The changes in the main lines of the RS with increasing dose of irradiation $D$ are as expected: in the non-irradiated samples (0), the intensity of the "defective" D-line is small ($\alpha = 0.15$) which means that the initial film has a not bad quality though not being perfect. It looks natural for the large size polycrystalline films. Increasing irradiation causes the D-line to increase and then broaden and decrease, while the 2D-line monotonically decreases and disappears which shows disappearance of the intact areas of graphene lattice. Together with the D-line, new defect lines appear: D'- line (1620 cm$^{-1}$) and (D+G)-line (2950 cm$^{-1}$). As to the G-line, it remains approximately constant and only broadened because of the appearance of the nearly located D'-line. Further increase of the irradiation leads to decrease and broadening of D-line, so the ratio $\alpha$ decreases. On the latter stages, the G-line also broadened and decreases. Eventually, all RS structure disappears at high level of disorder which can be explained by the film destruction.

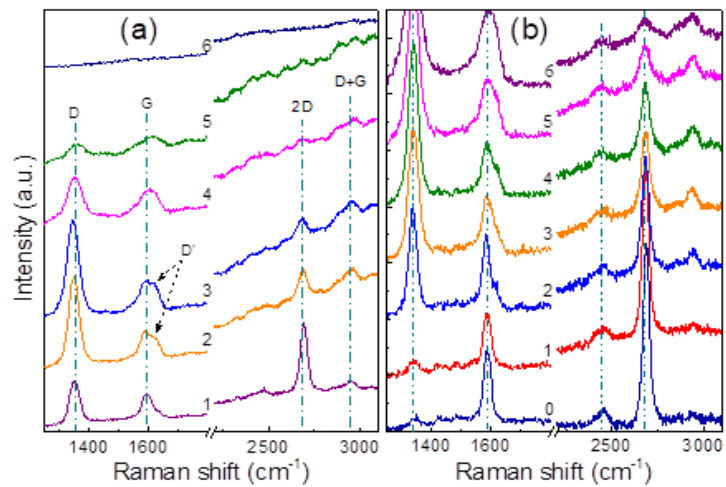

**Figure 3.** Raman scattering spectra for two series of samples irradiated with $C^+$ ions (a) and $Xe^+$ ions (b). For C-series, irradiation dose $D$ in units of $10^{14}$ cm$^{-2}$: 1–0 (non-irradiated, but after electron-beam lithography), 2–0.5, 3–1.0, 4–2.0, 5–10; for Xe-series, $D$ in units of $10^{13}$ cm$^{-2}$: 0–0 (initial), 1–0.15, 2–0.3, 3–0.5, 4–1.0, 5–2.0, 6–4.0.

Concentration of structural defects introduced by impact particles depends on their energy $E$ and mass $M$. Therefore, to compare the results of irradiation with different ions, one has to replace the irradiation dose $D$ by another parameter proportional to $D$ – concentration of introduced structural defects $N_D = kD$, with numerical coefficient $k$ dependent on mass and energy of the incident ion. The value of $k$ reflects the average fraction of carbon vacancies in the graphene lattice per ion impact.

A STRIM computer simulation of this process was performed in Ref. [23]. On the basis of this simulation, we plot in Fig. 4 the values of $k$ as a function of $M$ for different $E$. It follows from Fig.4 that for C ions ($M = 12$) with $E = 35$ keV, $k \approx 0.08$, whereas for Xe ($M = 131$) with the same energy, $k$ is 10 times more ($k \approx 0.8$). Therefore, to equalize the density of the introduced defects $N_D$, we had to use for Xe ions one order of magnitude smaller doses than for C ions. The dependences of $\alpha = I_D/I_G$ and $\beta = I_{2D}/I_G$ as a function of $N_D$ for both series of samples are shown in Fig. 5. One can see that in this scale, the experimental data for the C- and Xe-series almost coincide. This shows that indeed, the degree of disorder can be characterized by the concentration of introduced defects $N_D$.

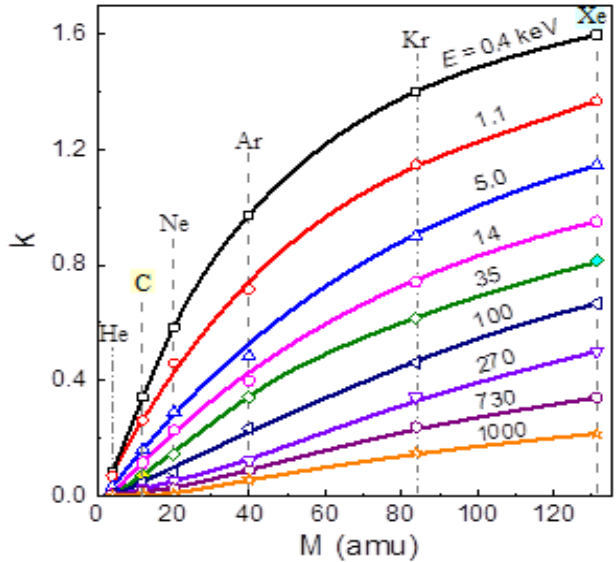

**Figure 4.** Dependence of $k$ on the ion mass $M$ for different energies $E$ shown in keV. Masses of some ions are indicated by vertical dotted lines.

Non-monotonic behavior of α = $I_D/I_G$ with increase of $N_D$ (or with decrease of the mean distance between defects $L_D = N_D^{-1/2}$) is described by the empirical model developed in Ref. [24]. In this model, a single defect causes modification of two length scale $r_A$ and $r_S$ ($r_A > r_S$), (inset in Fig. 6).

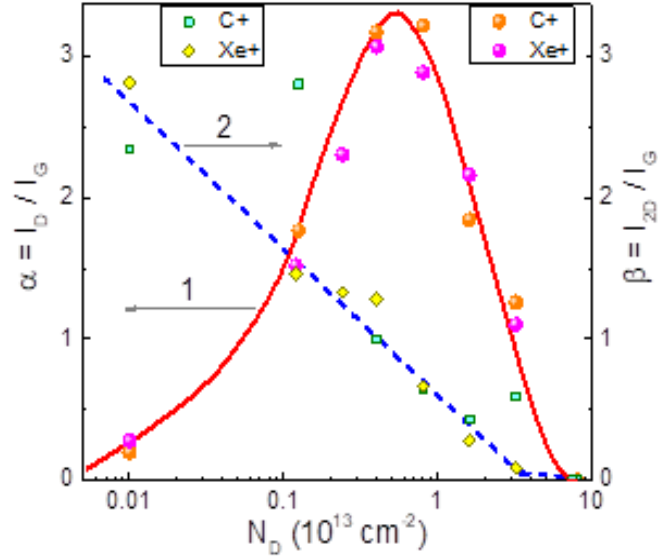

**Figure 5.** Normalized amplitude of the D-line α = $I_D/I_G$ (1) and the 2D-line β = $I_{2D}/I_G$ (2) for samples irradiated with $C^+$ ions and $Xe^+$ ions as a function of the density of introduced defects $N_D$.

Just in the near vicinity of the defect, the area $S = \pi r_S^2$, shown in red, is structurally disordered, while in the region $r_S < r < r_A$, shown in green, the lattice structure is saved, though the proximity to a defect leads to breaking of the selection rules which allows the emission of D-line. The "activated" area responsible for the D-line is $A = \pi(r_A^2 - r_S^2)$. In the low-defect-density regime, the intensity of the D-line is linearly increases with $N_D$ which means that α = $I_D/I_G \sim L_D^{-2}$. The maximal value of α is achieved when $L_D$ decreases down to $r_A$ ($L_D \approx r_A$). Further decrease of $L_D$ leads to overlap between $A$- and $S$- areas which results in decrease of α. Within this model, the final equation for the dependence $α(L_D)$ has the form [17]:

$$\frac{I_D}{I_G} = C_A \exp\left(-\frac{\pi r_S^2}{L_D^2}\right)\left[1 - \exp\left(-\frac{\pi(r_A^2 - r_S^2)}{L_D^2}\right)\right] + C_S\left[1 - \exp\left(-\frac{\pi r_S^2}{L_D^2}\right)\right] \quad (1)$$

Figure 6 shows the result of fitting the theoretical curve, Eq. (1), with experimental data. The best fit was achieved with $C_A$ =5.4, $C_S = 0$, $r_S = 1.55$ nm, and $r_A = 4.1$ nm. One can see from Fig. 6 that the maximum value $I_D/I_G$ occurs at $L_D = 5$ nm which is indeed close to $r_A$ (4.1 nm).

In other words, at this $N_D$, the "activated" A-area fills all the sample space. The value α for initial, non-irradiated sample was placed "by hand" on the curve. This allows one to estimate $L_D$ for these samples and, therefore, the density of EBL induced defects in the initial film as $N_D(0) \approx 8 \times 10^{10}$ cm$^{-2}$.

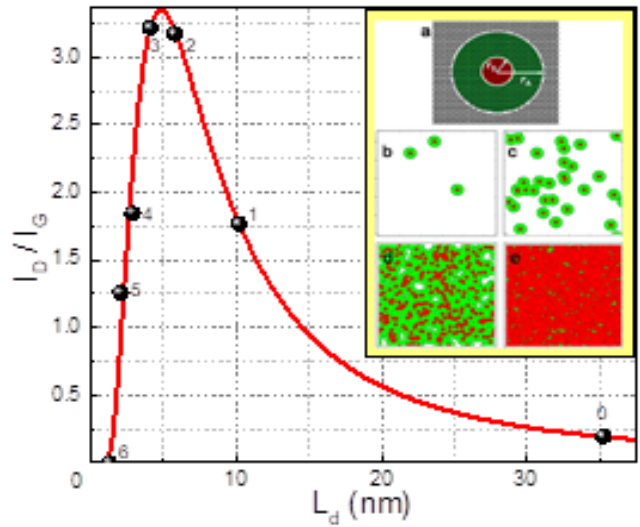

**Figure 6.** Ratio $I_D/I_G$ as a function of the mean distance between defects $L_D$. Solid line represents Eq. (1) with $C_A = 5.4$, $C_S = 0$, $r_S = 1.55$ nm, and $r_A = 4.1$ nm. Inset shows schematics of a single defect (a) and an irradiated sample with increase of defect concentration $N_D$ (b)-(e) [24].

## IV. Sample resistivity

In Fig. 7, the resistivity $R$ at room temperature for all samples is shown together with transformation of α. One can see that increase of $N_D$ leads to the strong continuous increase of $R$ over many orders of magnitude. Figure 7 shows also that the maximum α= $I_D/I_G$ corresponds to the film resistivity $R_m \approx 20$ kOhm, which is approximately equal to the quantum of two-dimensional resistance $h/e^2 \approx 25.8$ kOhm. Measurements of the current-voltage characteristics (*I-V*) show that for the highly disordered samples 5 and 6, *I-V* is strongly non-linear even at very small current. This allows us to conclude that these strongly irradiated samples with $R \gg R_m$ are more amorphous rather than crystalline, which agrees with the fact, that for $N_D > 10^{13}$ cm$^{-2}$, the 2D-line in the RS disappears (Fig. 3). That is why the temperature dependences of resistance $R(T)$ were measured only for samples 0 – 4 in order to remain in the crystalline structure and in the ohmic regime.

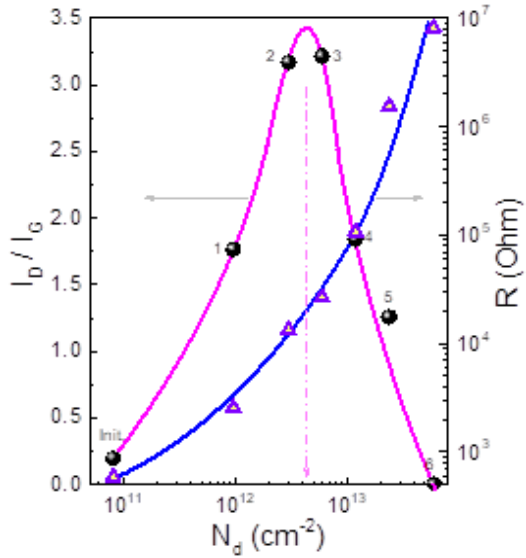

**Figure 7.** Ratio $I_D/I_G$ and the sample resistance $R$ at 300 K plotted as a function of the defect concentration $N_D$.

Figure 8 shows the temperature dependences $R(T)$ for all samples plotted on double logarithmic scale. It is seen, that sample 0 displays typical metallic behavior with $dR/dT > 0$, when $R$ slightly decreases with decrease of $T$. For sample 1, $R$ slightly increases with decreasing $T$ ($dR/dT < 0$), and finally, for samples 3 and 4, $R(T)$ increases exponentially, which is characteristic for strongly localized charge carriers. Figure 8 shows therefore gradual localization of charge carriers and metal-insulator transition in monolayer graphene. Let's discuss the transformation of the mechanisms of conductivity with increasing the irradiation-induced density of defects $N_D$.

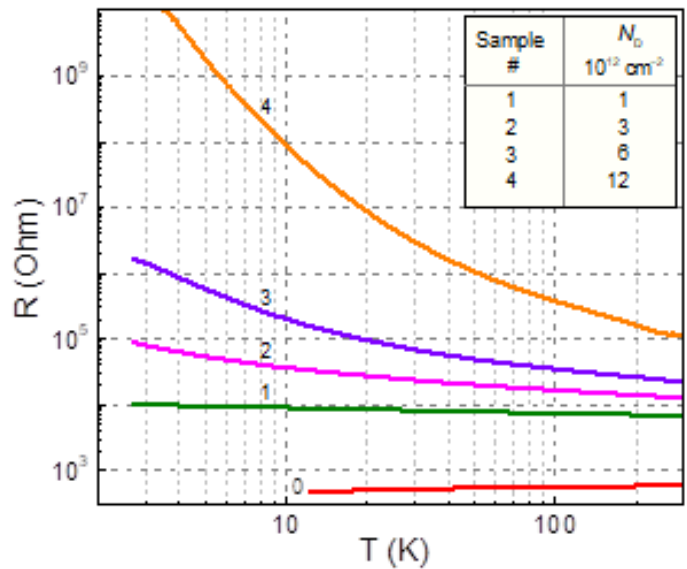

**Figure 8.** Resistivity of disordered monolayer graphene samples as a function of temperature plotted on the double logarithmic scale. Inset shows the density of structural defects in the samples.

1. **Weak Localization**

For sample 1, plot of the temperature dependence of resistance on the scale $R$ vs. log $T$ shows the logarithmic temperature behavior of $R$ with tendency to saturation at low temperatures (Fig. 9) which is characteristic for quantum corrections to the conductivity due to the regime of "weak localization" (WL) [25]. The WL regime is observed in disordered electronic systems at low temperatures when the electron motion is diffusive rather than ballistic. The WL correction come mostly from quantum interference between self-crossing paths in which an electron can propagate in the clock-wise and counter-clockwise direction around a loop. Due to the identical length of the two paths along a loop, the quantum phases cancel each other exactly. The two paths along any loop interfere constructively which leads to a higher net resistivity. Perpendicular magnetic field $B$ breaks this constructive interference which leads to decrease of the resistivity or increase of conductivity $\Delta\sigma = \sigma(B) - \sigma(0) > 0$.

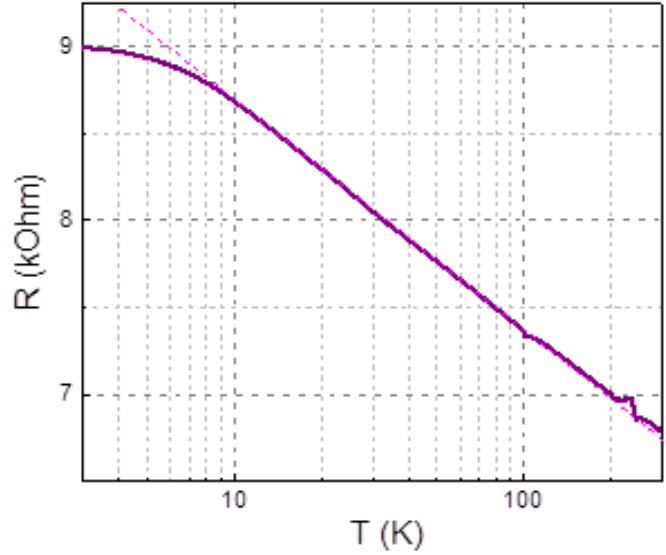

**Figure 9.** Resistivity of sample 1 from Fig. 8 as a function of log $T$.

WL regime of conductivity in monolayer graphene has important features due to the facts that charge carriers are chiral Dirac fermions, which are reside in two inequivalent valleys at the $K$ and $K'$ points of the Brillouin zone (see Fig. 1). Due to chirality, Dirac fermion acquires a phase of $\pi$ upon intra-valley scattering, which leads to destructive interference with its time-reversed counterpart and weak antilocalization (WAL). In this case, perpendicular magnetic field leads to decrease of the conductivity, $\Delta\sigma = \sigma(B) - \sigma(0) < 0$. Inter-valley scattering leads to restoration of WL because fermions in $K$ and $K'$ valleys have opposite chirality.

Quantum corrections to the conductivity of graphene have been intensely studied theoretically [26-28]. It is predicted that at relatively high temperatures WAL corrections will dominate, while with decreasing $T$ the WL corrections will dominate. There were several experimental papers reporting logarithmic dependence of conductivity on temperature and magnetic field at low temperatures [28-33]. However, in our sample 1, the logarithmic dependence is observed in wide temperature interval, starting from 300 K, which gives an opportunity to check in a very detailed way the theoretical predictions. For the magneto-conductance (MC), the theory [26] predicts

$$\Delta\sigma(B,T) = \frac{e^2}{\pi h}\left[F\left(\frac{B}{B_\varphi}\right) - F\left(\frac{B}{B_\varphi + 2B_i}\right) - 2F\left(\frac{B}{B_\varphi + B_*}\right)\right] \quad (2)$$

$$F(z) = \ln(z) + \psi\left(\frac{1}{2} + \frac{1}{z}\right), \quad B_{\varphi,i,*} = \frac{\hbar c}{4De}\tau^{-1}_{\varphi,i,*}$$

where $\psi$ is the digamma function, $\tau_\phi$ is the coherence time, $\tau_i^{-1}$ – is the inter-valley scattering rate, $\tau_*^{-1}$ – is the combined scattering rate of intra-valley and inter-valley scattering.

The result of fitting Eq. (2) to experimental data for magnetoconductance of sample 1 at different temperatures is illustrated on Fig. 10. In the process of fitting we are able to extract all three parameters entering the equation, which are shown in inset in Fig. 11. It turns out, that the parameters are temperature-dependent which is not predicted by theory. To prove the correctness of the fitting, we calculate conductance at zero magnetic field on the base of the obtained parameters. For calculation, we use the Equation (10) of [26] which can be rewritten in the form

$$\Delta\sigma(B=0,T) = -\frac{e^2}{\pi h}\left[\ln\left(1 + 2\frac{B_i}{B_\varphi}\right) + 2\ln\left(1 + \frac{B_*}{B_\varphi}\right) + 2\ln\left(\frac{B_\varphi}{1\text{T}}\right)\right] + A \quad (3)$$

where $A$ is a numerical constant, dependent upon the unit of magnetic field (chosen as 1T). The comparison of calculation Eq. (3) with experimentally measured conductivity $\sigma(T)$ is presented in Fig. 11. The good agreement proves the correctness of the chosen parameters. The saturation of $\tau_\varphi$ at low temperatures is well known in classical two-dimensional (2d) systems and may be connected with the existence of dephasing centers (for example, magnetic impurities) [34].

Obtained values of $B_\varphi$ allow us to determine the dephasing length $L_\varphi = (D\tau_\varphi)^{1/2} = (\hbar/4B_\varphi e)^{1/2}$. When the temperature decreases from 300 K to 3 K, $L_\varphi$ increases from 7 nm to 70 nm and then saturates. The maximal value of $L_\varphi$ allows us to estimate the density of dephasing centers as $2\times 10^{10}$ cm$^{-2}$.

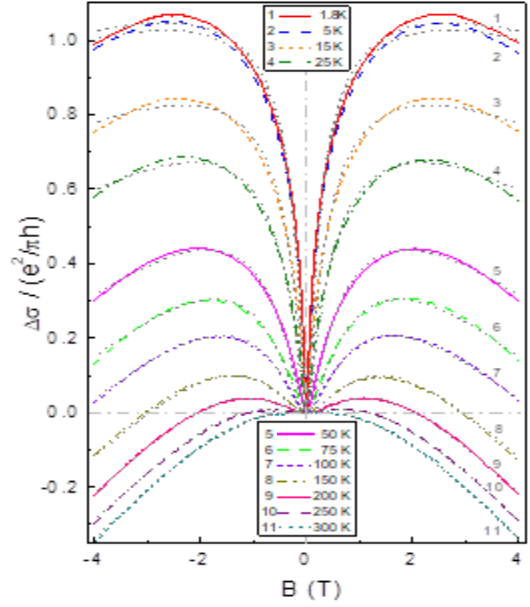

**Figure 10.** Magneto-conductance of sample 1 as a function of magnetic field at different temperatures; solid lines - experiment, dashed lines - Equation (2) with fitted parameters.

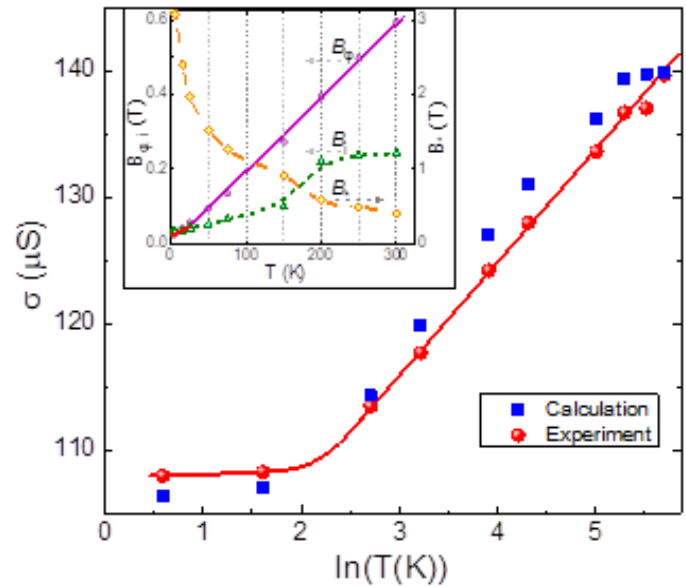

**Figure 11.** Conductivity of sample 1 as function of ln $T$. Circles present experimental data, squares present Equation (3) with parameters determined from fitting the magnetoconductance

(the constant A was chosen to be 11.7 $e^2/\pi h$). Inset shows the values of $B_\varphi$ (solid line), $B_i$ (short-dashed line) and $B_*$ (long-dashed line, right axis).

## 2. Strong localization

Let's discuss now samples 2 - 4 with pronounced insulating behavior. Plotting the data on the Arrhenius scale log $R$ vs. $1/T$ shows that the energy of activation continuously decreases with decreasing $T$ which is characteristic for the variable-range-hopping (VRH) conductivity [5]. There are two regimes of VRH depending on the structure of the density-of-states (DOS) $g(\varepsilon)$ in the vicinity of the Fermi level (FL) $\mu$. When $g(\varepsilon) = g(\mu) = $ const, $R(T)$ in the case of two-dimensional conductivity is described by the Mott (or "$T^{-1/3}$") law

$$R(T) = R_0 \exp(T_M)^{1/3} R(T) = R_0 ; \qquad R(T) = R_0 \exp(T_M)^{1/3} T_M = C_M[g(\mu)a^2]^{-1} \qquad (4)$$

Here $C_M = 13.8$ is the numerical coefficient [5], $a$ is the radius of localization.

The Coulomb interaction between localized carriers leads to appearance of the soft Coulomb gap in the vicinity of FL which in the case of 2d has a linear form:

$$g(\varepsilon) = |g - \mu|(\varepsilon^2/\kappa)^{-2} \qquad (5)$$

Here $\kappa$ is the dielectric constant. This leads to the Efros-Shklovskii (ES) (or "$T^{-1/2}$") law:

$$R(T) = R_0 \exp(T_{ES}/T)^{1/2}; \qquad T_{ES} = C_{ES} (e^2/\kappa a) \qquad (6)$$

where the numerical coefficient $C_{ES} = 2.8$ [5].

Coulomb interaction can alter the DOS only near the FL. Far from FL, the DOS is restored to its initial value, which is approximately equal to $g(\mu)$, see inset in Fig. 12. Denoting the half-width of the Coulomb gap as $\Delta$, one can conclude, therefore, that at $T \ll \Delta$, ES law has to be observed, while in the opposite case ($T \gg \Delta$), the Mott law should dominate.

There are a number of reports about observation of either Mott or ES laws in different disordered graphene based materials [35-38]. We show that in samples 3 and 4, both VRH laws are observable at different temperatures. In Fig. 12, log $R$ is plotted versus $T^{-1/3}$. At high temperatures, dependencies $R(T)$ are straightened on the scale $T^{-1/3}$, while at low temperatures they deviate to stronger dependence $\sim T^{-1/2}$. The latter shows the approach to the ES law which should be observed at the lowest temperatures. These plots allow us to determine both parameters $T_M$ and $T_{ES}$ (Table 1) and calculate the temperature $T_c$ of the crossover from $T^{-1/3}$ law to $T^{-1/2}$ law in the following way [39].

In VRH, only localized states in an optimal band of width $\varepsilon(T)$ near the Fermi level are involved in the hopping process. The band becomes continuously narrower with decreasing $T$. Hopping resistance is determined by the critical parameter $\xi_c$ [5]:

$$R(T) = R_0 \exp \xi_c; \qquad \xi_c = (2r/a) + (\varepsilon/T) \qquad (7)$$

Here energy and temperature are measured in the same units and $r$ is the mean distance of hopping. In the Mott regime, $g(\varepsilon) = g(\mu) = $ const. and, therefore, the total number of states in the optimal band is $N(T) = g(\mu)\varepsilon(T)$ and the mean distance between states in two dimensions is $r \approx [g(\mu)\varepsilon]^{-1/2}$. Substituting in Equation (7), one can find from the minimal value of $\xi_c$ determined from $d\xi_c/d\varepsilon = 0$:

$$\varepsilon(T) = T^{2/3}[g(\mu)a^2]^{-1/3} \qquad (8)$$

This gives the relationship between $T$ and the width of the optimal band: $T = [g(\mu)a^2]^{1/2}\varepsilon^{3/2}$. At the crossover temperature $T_c$ we get $\varepsilon = \Delta$, so

$$T_c = [g(\mu)a^2]^{1/2} \Delta^{3/2} \qquad (9)$$

Being inside the Coulomb gap, the crossover temperature can be determined from $g(\varepsilon = \Delta) = g(\mu)$, which gives $\Delta = g(\mu)(e^2/\kappa)^2$. Substituting into Equation (9) and using expressions for $T_M$ and $T_{ES}$ from Equation (4) and Equation (6), we get

$$T_c = (C_M{}^2/C_{ES}{}^3)(T_{ES}{}^3/T_M{}^2) \approx 8.6 \, (T_{ES}{}^3/T_M{}^2) \qquad (10)$$

Table 1.

| S # | $T_M$(K) | $T_{ES}$(K) | $T_c$(K) | $\Delta$(K) | $a$(nm) |
|---|---|---|---|---|---|
| 3 | 308 ± 54 | 50 ± 7 | 11.4 ± 0.8 | 12 ± 1.2 | 385 ± 54 |
| 4 | 5962 ± 305 | 490 ± 12 | 29 ± 5 | 60 ± 6 | 39 ± 1 |

The values of $T_c$ calculated from Eq. (10) for samples 3 and 4 are given in Table 1 and shown as arrows in Fig. 12. The good agreement shows the correctness of the obtained numerical coefficient. The functional proportionality of $T_c$ to the ratio $T_{ES}{}^3/T_M{}^2$ has been obtained earlier [40, 41], but with significantly different numerical coefficient. The latter is, however, crucial for comparison with experiment. We can also estimate the width of the Coulomb gap $\Delta = (C_M/C_{ES}^2)(T_{ES}^2/T_M)$ which is shown in Table 1. One can see, that the ES law is observed, indeed, at $T < \Delta$.

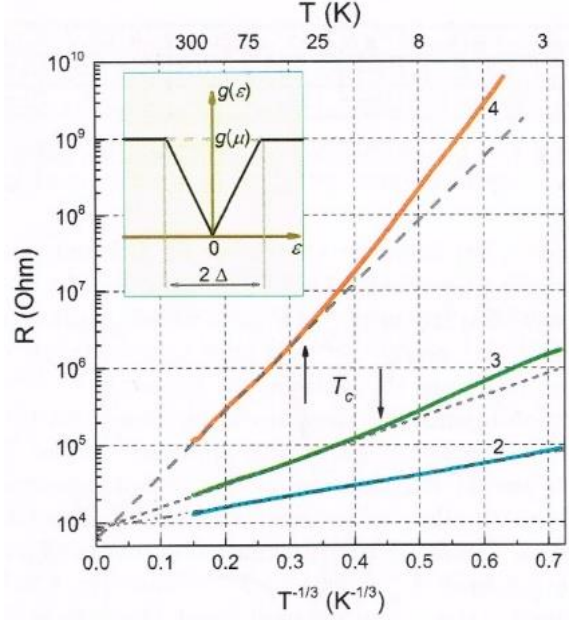

**Figure 12.** Resistivity of samples 2-4 from Fig. 8 plotted as log $R$ vs. $T^{-1/3}$. Arrows indicate the temperature $T_c$ of crossover from the Mott ("$T^{-1/3}$- law ") to the ES ("$T^{-1/2}$- law ") mechanism of VRH. Inset shows schematics of two-dimensional Coulomb gap in the density-of-states in the vicinity of the Fermi level.

### 3. Hopping magnetoresistance

In this subsection, the results of measurements of magnetoresistance (MR) in samples 2-4 with hopping mechanism of conductivity are presented and discussed. Measurements of MR in samples 2–4 were performed at temperatures down to 1.8 K and in magnetic fields up to $B = 8$ T in perpendicular $B_\perp$ and in-plane(parallel) $B_\parallel$ geometry. It was observed that $B_\perp$ leads to negative MR (NMR) while $B_\parallel$ results in positive MR (PMR) at low temperatures, Fig. 13. This anisotropy shows unambiguously that MR in perpendicular and parallel fields has a different origin: NMR is determined by the orbital mechanisms, while PMR is determined by the spin polarization. In that order we will discuss the results of measurements.

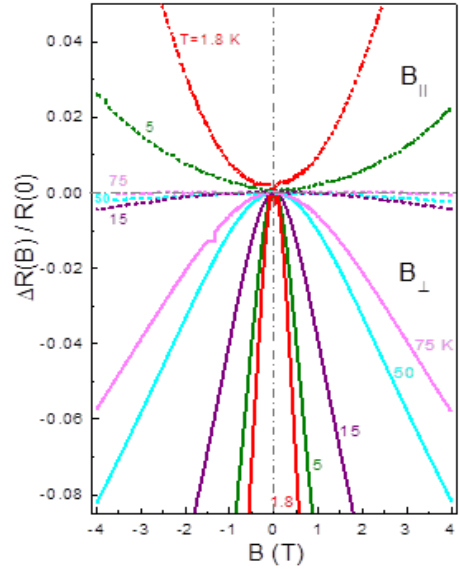

**Figure 13.** Magnetoresistance $\Delta R(B)/R(0)$ for sample 3 in perpendicular (solid lines) and parallel (dotted lines) magnetic fields $B$ at different temperatures in K, shown near each curve.

### 3.1. Negative MR in perpendicular magnetic fields

Figure 14 shows the NMR curves $\Delta R(B)/R(0) \equiv [R(B)-R(0)]/R(0)$ at different $T$ for all three samples on a linear scale (A) and on a double logarithmic scale (B). One can see, that NMR at fixed $T$ decreases with an increase of disorder from sample 2 to 4. Temperature dependencies show that NMR increases with decreasing $T$ for samples 2 and 3. For sample 4, $\Delta R(B)/R(0)$ first increases with decreasing $T$, but at $T < 10$ K, NMR rapidly decreases and the curves seek to change the sign. It could be due to the standard PMR caused by the shrinkage of the wave functions in perpendicular magnetic fields [5]. Therefore, we will discuss the NMR for sample 4 only down to 10 K.

In Fig. 14B, NMR curves are plotted on the log–log scale. On this scale, the slope to the curve is equal to the power $m$ in $\Delta R/R \sim B^m$. Quadratic dependence ($m = 2$) is observed at low fields up to some value $B^*$. At $B > B^*$, the quadratic dependence is replaced by the linear one ($m = 1$) and then by sublinear dependencies. The values of $B^*$ are shown in Fig. 14B by arrows.

Very weak effect of NMR (about 1–2%) in VRH regime was earlier observed in three-dimensional (3d) conductivity in heavily doped and compensated Ge (for a review, see [42]). In 2d, a significant NMR in the VRH regime has been observed in perpendicular magnetic fields in different semiconductor systems [43,44]. Theoretically the effect of NMR in the VRH regime has been discussed in [45-47]. The main idea of the "orbital" model suggested by Nguen, Spivak and Shklovskii [45] is based on the following consideration.

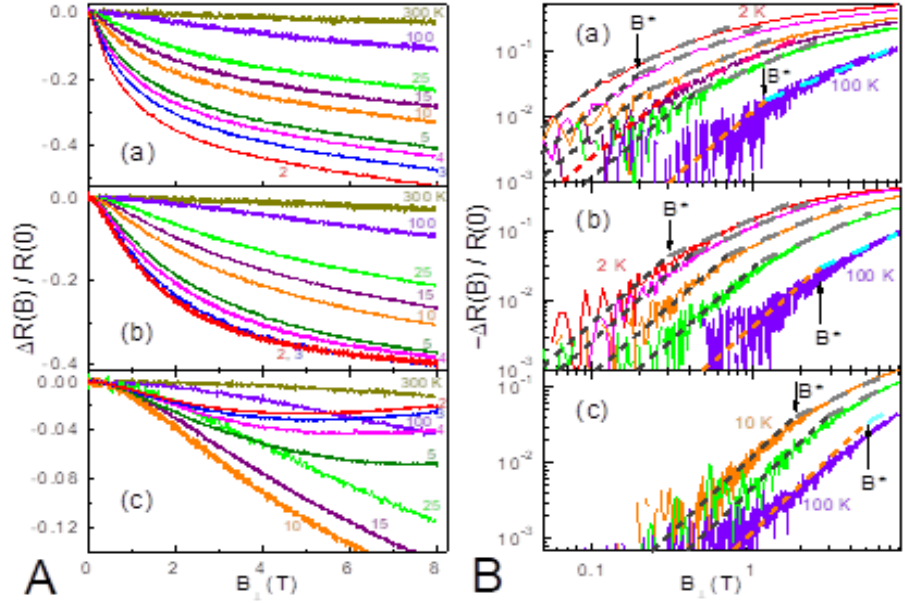

**Figure 14.** NMR of samples 2 (a), 3 (b) and 4 (c) at different temperatures (*T*, K) shown near the curves. The dependences are plotted on the linear scale (A) and on double logarithmic scale (B). The density of structural defects $N_D$ (cm$^{-2}$): sample 2 – 3 ×10$^{12}$, sample 3 – 6 ×10$^{12}$, sample 4 – 1.2 ×10$^{13}$. Short dashed lines correspond to $m = 2$ in $\Delta R/R \sim B^m$, dashed lines to $m = 1$. The arrows indicate $B^*$ - the end of quadratic dependence.

In VRH, only part of localized states with energy levels within the so-called "optimal band" around FL $\epsilon(T)$ is involved in the hopping process. In "Mott VRH", $\epsilon(T)$ decreases with a decrease of temperature [5]:

$$\epsilon(T) = T^{2/3}[g(\mu)\xi^2]^{-1/3} \qquad (11)$$

Correspondingly the hopping distance $r_h$ increases, which gives $r_h \sim T^{-1/3}$:

$$r_h \approx [g(\mu)\,\epsilon(T)]^{-1/2} \approx \xi(T_M/T)^{1/3} \qquad (12)$$

In "ES VRH", $r_h \sim T^{-1/2}$. As a result, in VRH regime at low *T*, $r_h$ becomes much larger than the mean distance between localized centers, and therefore, the probability of the long-distance hop becomes dependent on the interference of many paths of the tunneling through the intermediate sites which include a scattering process (Fig. 15). All these scattered waves, together with non-scattered direct wave, contribute additively to the amplitude of the wave function $\Psi_{12}$ which reflects the probability for a charge carrier localized on site 1 to appear on site 2. There is no backscattering, and scattered waves decay exponentially with increasing distance as $\exp(-2r/\xi)$, therefore only the shortest paths contribute to $\Psi_{12}$. All these paths are concentrated in a cigar-shaped domain of the length $r_h$, the width $D \approx (r_h \xi)^{1/2}$ and the area $A \approx \alpha r_h^{3/2} \xi^{1/2}$, where $\alpha \leq 1$ is a numerical coefficient.

After averaging over different configurations, the contribution of the scattered sites to the total hopping probability vanishes due to destructive interference. The perpendicular magnetic field suppresses the interference which leads to the increase of the hopping probability and, therefore, to NMR

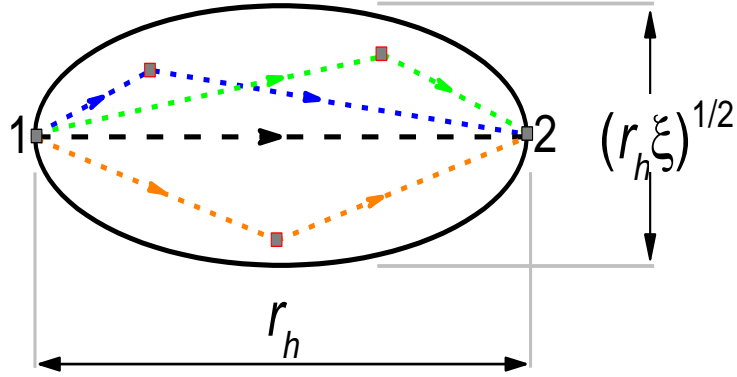

**Figure 15.** Schematics of the cigar-shaped region with localized states contributing to the probability of an electron tunneling from center 1 to center 2.

In accordance with theoretical considerations, NMR as a function of magnetic field $B$ has to be linear at moderate fields and quadratic at very low fields. In the "orbital" model, it is natural to normalize the magnetic field using the ratio $\eta = \Phi_B/\Phi_0$ where $\Phi_0 = h/2e \approx 2 \times 10^{-15}$ W is the magnetic flux quantum and $\Phi_B = B \cdot A$ is the magnetic flux through the cigar-shape area.

We suggest that $B^*$ corresponds to $\eta = 1$ which gives $B^* = \Phi_0/A \sim r_h^{-3/2} \xi^{-1/2}$. Taking into account Eq. (12), one gets ($\alpha \approx 1$):

$$B^* = \Phi_0 \xi^{-2} (T_M/T)^{-1/2} = \lambda T^{1/2}; \quad \lambda = \Phi_0 \xi^{-2} T_M^{-1/2} \qquad (13)$$

In Fig. 16, the values of $B^*$ for all samples are plotted as a function of $T^{1/2}$. One can see that, indeed, $B^* \sim T^{1/2}$. Coefficient $\lambda$ is equal to 0.1, 0.24 and 0.58 T·K$^{-1/2}$ for samples 2, 3 and 4 respectively. Knowledge of $\lambda$ allows us to estimate with accuracy of $\alpha$ the values of localization radius $\xi$. For samples 2, 3 and 4, with $T_M$ = 68, 308 and 5960 K [18], this gives $\xi$ = 50, 22 and 7 nm correspondingly, which looks quite reasonable: $\xi$ decreases with an increase of disorder (from sample 2 to sample 4).

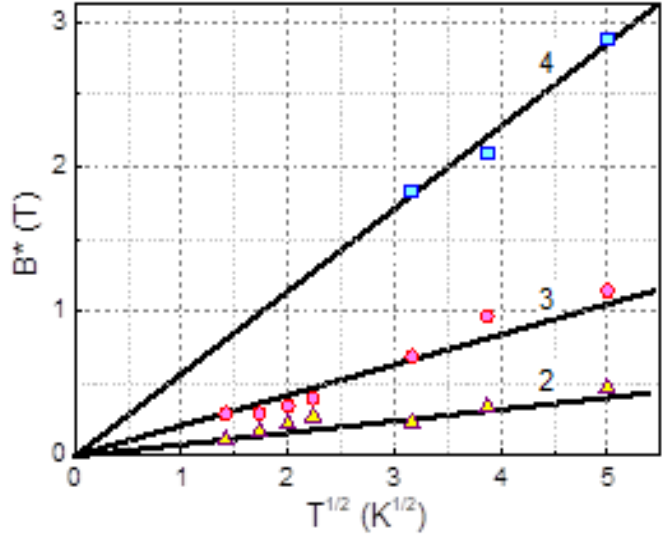

**Figure 16.** The values of $B^*$ as a function of $T^{1/2}$. The sample numbers are shown near the straight lines.

We also use the values of $B^*$ in an attempt to merge the NMR data for all samples and all temperatures below 25K. In Fig. 17, NMR curves from Fig. 14A are plotted as a function of dimensionless parameter $B/B^*$. One can see that all curves are merged in a universal dependence. No saturation of NMR is observed up to $\eta > 60$.

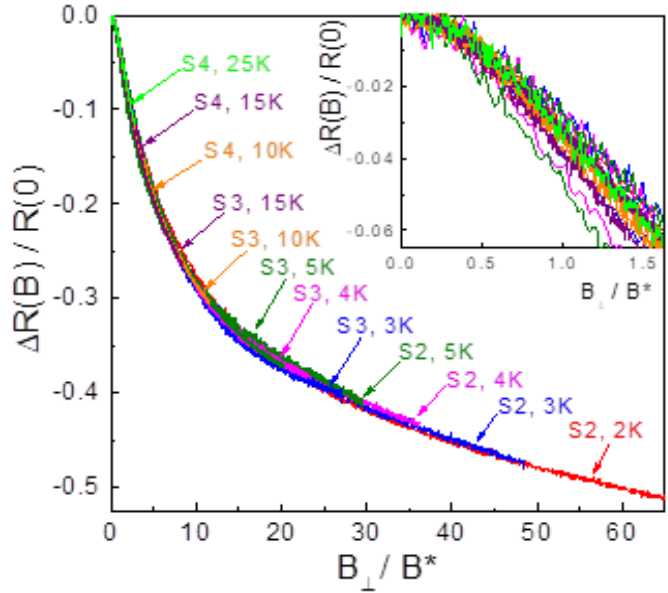

**Figure 17.** The NMR data for different samples and different temperatures plotted as a function of dimensionless magnetic field $B_\perp/B^*$. The arrows and numbers show the end of each curve and indicate the sample (2–4) and $T$. In inset, the NMR data are shown for small values $B_\perp/B$.

## 3.2. Positive MR in parallel magnetic fields

In parallel magnetic fields, PMR is observed at low temperatures, Fig. 13. Very small NMR at high temperatures could be explained as the traces of NMR due to the possible folds on the surface of monolayer graphene film, where the parallel magnetic field has a perpendicular influence component. PMR appears only at low temperatures, at $T < 5$ K and increases with decreasing $T$. PMR is proportional to $B_\parallel^2$ at low fields, and becomes linear with increasing $B$ (Fig. 18).

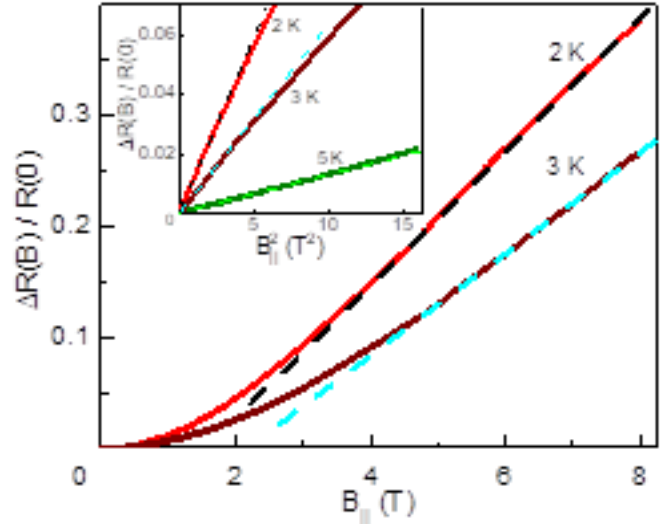

**Figure 18.** PMR of sample 3 plotted as a function of $B_\parallel$. In inset, PMR data for low fields are plotted on a quadratic scale.

PMR in parallel magnetic fields has been observed earlier in 2d VRH regime in different systems: in $Al_xIn_{1-x}Sb/InSb$ quantum well [43], in a $GaAs/Al_xGa_{1-x}As$ heterostructure [48]. Because the parallel magnetic field couples only to the electron spin, it means that the spin state of localized electrons influences the hopping conductivity despite the fact that it is not included explicitly in the expressions for VRH. The explanatory models of this effect are based on two ideas. The authors [49] considered the case when intermediate scattering centers (see Fig. 15) should be occupied to produce a negative scattering amplitude. Thus, the interference is depended on the mutual spin orientation of the hopping electron and electron localized on scattering center. In magnetic fields, all localized spins are aligned which increases the destructive interference and results in an increase of the resistance. Another mechanism was suggested in Refs. [50] and [51]. In this model, it is recognized that a certain fraction of the states can accommodate two electrons. Double occupancy is possible if the on-site Coulomb repulsion, $U$, between the electrons is smaller than the width of the energy distribution function of the localized states. It was already mentioned that in VRH, only localized states with energy level within the narrow optimal band of width ε(T) around μ are involved in the hopping process, Eq. (11). However, for some states, which cannot participate in VRH at given temperature because the energy of the first electron $\varepsilon^{(1)}$ is well below μ, the energy of the second electron $\varepsilon^{(2)} = \varepsilon^{(1)} + U$ may be located just within the optimal band, Fig. 19. This allows those states to participate in the VRH at zero magnetic field. In strong field limit, all spins are aligned along the field and, therefore, double occupation become impossible which leads to increase of the hopping resistivity.

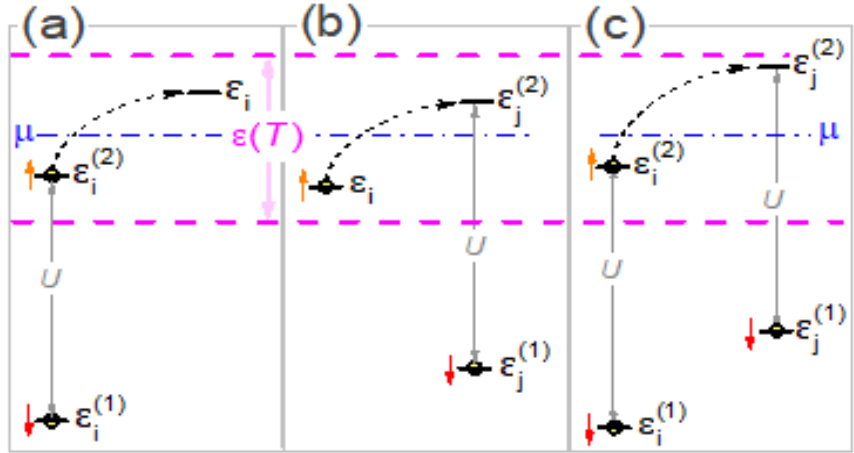

**Figure 19.** Schematic representation of the possible hopping motion via the double occupied states where the double occupied state is an initial one $\varepsilon_i$ (a), final $\varepsilon_j$ (b) and both initial and final states (c). μ represents the position of the Fermi level, dashed lines show the width ε(T) of the optimal band at given temperature.

Figure 13 shows that PMR is observed in our samples only at low temperatures. We believe that this fact supports the mechanism [50] based on participation of the double occupied states, because the mechanism suggested in Ref. [49] has no limitation for observation at all temperatures. In the mechanism [50], however, contribution of the double occupied states in VRH is important only at low T, when the width of the optimal band ε(T) which decreases with decreasing T, becomes less than U, otherwise in the case of an opposite inequality, $U \ll \varepsilon(T)$, the localized states will either participate or not participate in VRH independently of the existence of the double occupied states. At moderate magnetic fields, theory [51] predicts the linear dependence $\Delta R/R \sim (g_L \cdot \mu_B \cdot B)/T$, where $g_L$ is the Lande-factor and $\mu_B$ is the Bohr magneton, while at weak fields one expects the quadratic dependence $\Delta R/R \sim B^2$. This agrees with experiment (Fig. 18). Theory predicts also saturation PMR at strong fields when all electron spins are polarized. In our samples, no tendency to saturation was observed in magnetic fields up to 8T.

## V.     Conclusions

Gradual localization of charge carriers and the metal-insulator transition was realized and studied in a series of monolayer graphene (MG) samples disordered by means of the ion irradiation. Degree of disorder was controlled using measurements of the Raman scattering spectra. It was shown that disorder introduced by irradiation with different dose $D$ of different ions can be unified by using such a parameter as the density of structural defects $N_D = kD$, where coefficient $k$ depends on the mass and energy of the impact ion.

For slightly disordered MG sample, measurements of the temperature dependence of conductivity and magnetoresistance show that the mechanism of conductivity is determined by the regime of weak localization and antilocalization due to chirality of charge carriers in MG. Further increase of disorder leads to strong localization regime, when the conductivity is described by the variable-range-hopping (VRH) mechanism. A transition from the "Mott regime" to the "Efros-Shklovskii regime" of VRH was observed with decreasing temperature.

Magnetoresistance (MR) of strongly disordered samples with VRH mechanism of conductivity showed different behavior: in perpendicular magnetic fields resistance decreases (negative MR, NMR), while parallel magnetic fields lead to positive MR (PMR) at low temperatures. The NMR effect is explained by the "orbital" mechanism based on the interference of many paths through the intermediate sites in the probability of the long-distance tunneling in the VRH regime. The PMR effect in parallel fields was explained by suppression the hopping transitions via double occupied states due to spin polarization in strong magnetic fields.

Localization radius of charge carriers for samples with different degree of disorder was estimated.

## VI.    Acknowledgements

We are thankful to D. Naveh, A. Haran, L. Wolfson, T. Havdala who took part in various stages of the measurements.  This work was supported by the Jack and Perl Resnick Institute of Advanced Technology research foundation.